\documentclass[aoas]{imsart}

\RequirePackage{amsthm,amsmath,amsfonts,amssymb}
\RequirePackage[authoryear]{natbib}
\RequirePackage[colorlinks,citecolor=blue,urlcolor=blue]{hyperref}
\RequirePackage{graphicx}
\usepackage{bm}
\usepackage{float}
\usepackage{makecell}
\usepackage{wrapfig}

\startlocaldefs
\theoremstyle{plain}

\theoremstyle{remark}

\endlocaldefs

\begin{document}

\begin{frontmatter}
\title{Functional Accelerated Failure Time Models for Predicting Time Since Cannabis Use}
\runtitle{Functional Accelerated Failure Time Models}

\begin{aug}
\author[A]{\fnms{Weijia}~\snm{Qian}\ead[label=e1]{weijia.qian@emory.edu}\orcid{0009-0001-1600-5813}},
\author[B]{\fnms{Erjia}~\snm{Cui}\ead[label=e2]{ecui@umn.edu}},
\author[C]{\fnms{Ashley}~\snm{Brooks-Russell}\ead[label=e3]{ashley.brooks-russell@cuanschutz.edu}}
\and
\author[A]{\fnms{Julia}~\snm{Wrobel}\ead[label=e4]{julia.wrobel@emory.edu}}
\address[A]{Department of Biostatistics and Bioinformatics,
Emory University \printead[presep={ ,\ }]{e1,e4}}
\address[B]{Division of Biostatistics and Health Data Science, University of Minnesota \printead[presep={,\ }]{e2}}
\address[C]{Injury and Violence Prevention Center, University of Colorado Anschutz Medical Campus \printead[presep={,\ }]{e3}}
\end{aug}

\begin{abstract}
Cannabis consumption impairs key driving skills and increases crash risk, yet few objective, validated tools exists to identify acute cannabis use or impairment in traffic safety settings. Pupil response to light has emerged as a promising biomarker of recent cannabis use, but its predictive utility remains underexplored. We propose two functional accelerated failure time (AFT) models for predicting time since cannabis use from pupil light response curves. The linear functional AFT (\textit{lfAFT}) model provides a simple and interpretable framework that summarizes the overall contribution of a functional covariate to time-since-smoking, while the additive functional AFT (\textit{afAFT}) model generalizes this structure by allowing effects to vary flexibly with both magnitude and location of the functional covariate. Estimation is computationally efficient and straightforward to implement. Simulation studies show that the proposed methods achieve strong estimation accuracy and predictive performance across various scenarios and remain robust to moderate model misspecification. Application to pupillometry data from the Colorado Cannabis \& Driving Study demonstrates that pupil light response curves contain meaningful predictive signal, underscoring the potential of these models for traffic safety and broader biomedical applications.
\end{abstract}

\begin{keyword}
\kwd{functional data analysis}
\kwd{accelerated failure time model}
\kwd{survival analysis}
\kwd{cannabis impairment}
\end{keyword}

\end{frontmatter}

\section{Introduction}

Cannabis is legal for recreational or medical use in 48 states and the District of Columbia as of June 2024 \citep{NCSL2024CannabisOverview}. An estimated 64.2 million Americans aged 12 or older used cannabis in 2024 and that number has continued to rise \citep{samhsa2025nsduh}. With growing legalization and popularity, cannabis-impaired driving has become a pressing public safety concern. The National Transportation Safety Board (NTSB) reports that alcohol and cannabis are the two most frequently detected substances among fatally injured drivers and drivers arrested for impaired driving \citep{ntsb2022srr2202}. Acute cannabis use impairs multiple aspects of driving performance and driving-related cognitive skills, and is linked to an increased crash risk \citep{mccartney2021determining, rogeberg2019meta}. Despite these concerns, few objective and validated tools exist for detecting recent cannabis use, leaving law enforcement with limited means to ensure roadway safety \citep{spindle2021assessment, kosnett2023blood}. 

Recent studies have identified pupil response to light as a potential biomarker of recent cannabis use that may be swiftly and objectively measured by law enforcement at the roadside using wearable or handheld pupillometers \citep{steinhart2023video, godbole2024study, brooks2025pupillary}. In a standard pupil light response test, a brief light stimulus directed into the eye causes the pupil to constrict to a minimum diameter, after which it gradually dilates back toward its baseline size. \cite{steinhart2023video} reported that acute cannabis use is associated with reduced pupil constriction and slower rebound dilation during the light response test. \cite{brooks2025pupillary} and Bird et al. (2025) further showed that percent change in pupil diameter, along with other factors, predicts recent cannabis use. \cite{godbole2024study} modeled trajectories of percent change in pupil diameter during the light response test, referred to here as \textit{pupil light response curves}, using functional regression models. They showed that analyzing the full trajectory, rather than relying on scalar summaries, improves predictions of recent cannabis use. They also observed similar cannabis effects in both daily and occasional users, suggesting that pupil light response is not strongly influenced by tolerance.

Despite these encouraging results, important gaps remain. First, with the exception of \cite{godbole2024study}, which considered functional logistic regression, most analyses have relied on scalar summaries of pupil dynamics, such as rebound dilation or minimum constriction, rather than leveraging the full information contained in pupil light response curves. Second, prior work has focused on the binary classification of recent cannabis use, whereas estimating time since last use would be more informative for traffic safety, given that impairment is greatest within the first two hours after consumption \citep{arkell2020effect, tank2019impact}.

The objective of this study is to address these gaps by predicting time since last cannabis use from pupil dynamics using a functional data analysis (FDA) framework \citep{crainiceanu2024functional, ramsay2005functional, kokoszka2017introduction}. FDA is well-suited for modeling pupil light response curves as it leverages information in the full temporal structure of the curves. To model how recently an individual has consumed cannabis, we define a time-to-event outcome where the ``survival'' time represents the elapsed time since last consumption, and the censoring indicator reflects whether the individual is known to have smoked recently; control participants, who abstained from cannabis during the study and for at least eight hours prior to data collection, are therefore treated as censored at eight hours.

Because our focus is on predicting time since last use, we adopt accelerated failure time (AFT) models, which directly model survival time, in contrast to Cox proportional hazards (CoxPH) models that focus on the hazard function. Specifically, we propose a class of functional parametric AFT models that accommodate flexible choices of both error distribution and the functional covariate effect.

The remainder of our paper is organized as follows. Section \ref{StudyDesign} introduces the motivating pupillometer data, and Section \ref{literature} provides a brief review of functional regression models for time-to-event outcomes. Section \ref{methods} introduces the proposed functional AFT models and the associated estimation procedures. Section \ref{simulations} presents a simulation study evaluating both statistical performance and computational efficiency. Section \ref{data analysis} provides a comprehensive analysis of the motivating data using the proposed methods, and Section \ref{discussion} concludes with a discussion.

\subsection{Pupil light response curves and time since cannabis use}
\label{StudyDesign}

In the motivating Colorado Cannabis \& Driving Study (CCDS), 127 participants were recruited into either the cannabis user group (N = 96) or the non-user group (N = 31) based on self-reported use history \citep{brooks2025impact}. Non-users were defined as not having consumed cannabis in the month prior to enrollment, and users were asked to abstain for at least 8 hours before study participation. At baseline, all participants completed a series of assessments including simulated driving performance, blood THC concentration, and pupil light response measured using the NeurOptics PLR-3000 pupillometer.

Following baseline measurements, users were instructed to smoke cannabis \emph{ad libitum} for up to 15 minutes, while non-users rested for an equivalent period. All baseline assessments, including pupil light response, were then repeated at approximately 40 minutes (median = 40 minutes, range = 27-56) and 100 minutes (median = 97 minutes, range = 85-120) after the start of cannabis inhalation. These two post-smoking assessments were chosen to represent peak impairment (40 minutes) and a time window more reflective of a roadside evaluation (100 minutes). The interval between cannabis inhalation and pupil response measurement varied across participants due to differences in data collection timing. We exploit this natural variability in study timing to understand how pupil light response curves can be used to predict time since most recent cannabis inhalation.

\begin{figure}[H]
\includegraphics[width=1\textwidth]{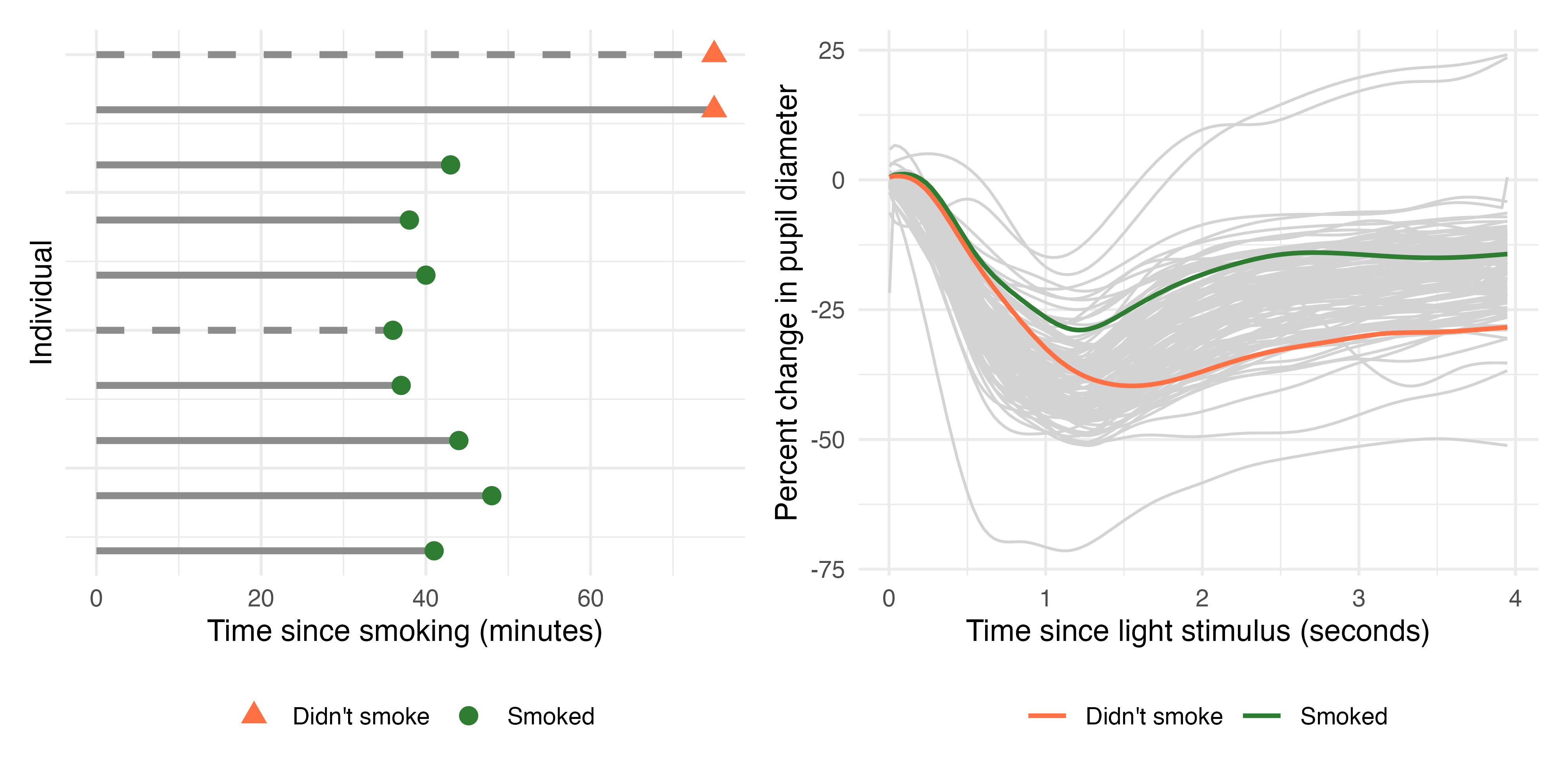}
\caption{Left: Pupil light response curves for the right eye with one representative subject from each use group. Right: Time since smoking for 10 sampled subjects, with dashed lines marking the two subjects shown on the left. Non-smokers are displayed as censored at 75 minutes for illustration.}
\label{fig_curves}
\end{figure}

The left panel of Figure~\ref{fig_curves} illustrates the distribution of time from smoking to the first assessment for 10 randomly sampled subjects, with dashed lines corresponding to the two highlighted subjects from the left panel. As shown, the interval between smoking and test administration varies across participants. For non-users, time since last cannabis use is treated as right-censored, since the exact time is unknown but is guaranteed to exceed a censoring threshold, given that no use occurred during the study period.

Let $X_i(s)$ denote the pupil response curve for the right eye at $s$ seconds since the light stimulus for participant $i$. The right panel of Figure~\ref{fig_curves} displays curves $X_i(s), i = 1,\ldots, 127$ at the first (approximately 40 minute) assessment, with one representative user and non-user highlighted. For a typical pupil response, the pupil constricts to its minimum diameter after the light stimulus at $s = 0$, then slowly rebounds towards its pre-stimulus size. The user (green line) shows reduced constriction following the light stimulus compared to the non-user (orange line). 

This study design motivates our development of a functional accelerated failure time (AFT) model, which uses pupil light response curves as functional predictors to estimate time since cannabis use as the outcome of interest.  We will model time since cannabis use at the 40 and 100 minute post-smoking assessments as separate outcomes.

\subsection{Time-to-event models with functional predictors}
\label{literature}

Analyzing time-to-event data with functional predictors is an active area of research. The Cox proportional hazards (CoxPH) model \citep{cox1972regression} is widely used in survival analysis and has been extended to incorporate functional covariates \citep{gellar2015cox, qu2016optimal, kong2018flcrm, cui2021additive}. The CoxPH model is semi-parametric, modeling the log hazard rather than modeling survival time directly, and leaving the baseline hazard unspecified while assuming covariates have linear effects on the log hazard ratio. 
However, when the proportional hazards assumption is violated, as often occurs in practice, the accelerated failure time (AFT) model offers a compelling alternative \citep{buckley1979linear, wei1992accelerated}. The AFT model directly estimates survival time from fitted values, making it particularly useful for roadside testing, where the goal is to estimate the probability that an individual has used cannabis within the past $\tau$ minutes (e.g., 60) given their pupil light response. Although the baseline survival distribution must be specified, AFT models are generally robust to distributional misspecification when estimating covariate effects \citep{hutton2002choice}. Compared to CoxPH models, AFT models also have straightforward interpretation analogous to linear regression, and estimate model parameters using the full maximum likelihood.

Only a few studies have extended AFT models to functional data. \cite{liu2025efficient} proposed a semi-parametric functional AFT model with an unspecified error distribution, estimated using a sieve maximum likelihood approach. \cite{ghosal2025functional} developed a functional time transformation model that assumes an unknown monotone transformation of survival time and a known error distribution, also estimated via a sieve method. While these approaches are flexible, they can be computationally intensive, and extrapolation outside the observed window of survival times may be less reliable without a parametric specification. In addition, existing work is limited to linear functional forms, which restricts the effect to be linear with the value of functional predictors at each point of the functional domain. Our work addresses these limitations by (1) proposing a fully parametric AFT model with functional predictors, which provides closed-form expressions for the hazard and survival functions, enabling straightforward extrapolation beyond the observed data range; (2) allowing the association between a functional predictor and survival time to vary nonlinearly across both the functional domain and the magnitude of the functional predictor; and (3) providing a ready-to-use R package facilitating broader adoption of our methods in applied research. To our knowledge, this is the first fully parametric functional AFT model, and the first AFT model that accommodates both linear and additive effects of functional covariates. 


\section{Methods}
\label{methods}
We propose two models for analyzing right-censored survival data with functional covariates: the additive functional accelerated failure time model (\textit{afAFT}) and its special case, the linear functional accelerated failure time model (\textit{lfAFT}).

\subsection{Additive functional accelerated failure time model}
The \textit{afAFT} model allows flexible, potentially nonlinear effects of functional covariates on survival outcomes. Let $T_i$ denote the survival time for subject $i$ and $C_i$ the censoring time, assumed independent of $T_i$ conditional on covariates. We observe $Y_i = \min(T_i, C_i)$ and the event indicator $\delta_i = I(T_i \le C_i)$. For each subject, let $\bm{Z}_i=\{Z_{i1}, \dots, Z_{id}\}^\mathsf{T}$ denote a vector of scalar covariates, and $X_i(s)$ a functional covariate defined on domain $\mathcal{S}$. The \textit{afAFT} model is given by

\begin{equation}
\log T_i=\bm{Z}_i^\mathsf{T}\bm{\gamma}+\int_\mathcal{S}F\left[s, X_i(s)\right]ds+\sigma\epsilon_i,
\label{eq:1}
\end{equation}

\noindent
where $\bm{\gamma}$ is the coefficient vector for the scalar covariates, $F(\cdot,\cdot)$ is an unknown smooth bivariate function, $\epsilon_i$ is an error term following a specified distribution (e.g., standard normal or logistic), and $\sigma$ is a scale parameter. This formulation allows the effect of $X_i(s)$ on $\log T_i$ to vary with both the functional location $s$ and the magnitude of $X_i(s)$. To ensure identifiability, we follow the literature and impose the constraint $E\{F[s, X_i(s)]\} = 0$ for each $s \in S$ \citep{cui2021additive}.

\subsection{Linear functional accelerated failure time model}
The \textit{lfAFT} model arises as a special case of the \textit{afAFT} when $F(s, x)$ is restricted to be linear in $x$, i.e., $F(s, x) = x\beta(s)$. The model can then be written as

\begin{equation}
\log T_i = \bm{Z}_i^\mathsf{T}\bm{\gamma} + \int_S X_i(s)\beta(s)ds + \sigma\epsilon_i.
\label{eq:2}
\end{equation}

\noindent
Interpretation in the \textit{lfAFT} model is analogous to scalar-on-function regression. $\beta(s)$ acts as a weight function summarizing how $X_i(s)$ contributes to the logarithm of survival time across the functional domain $\mathcal{S}$. Centering $X_i(s)$ is often helpful, as it allows $\beta(s)$ to be interpreted relative to a reference trajectory (e.g., the study population mean). The term $\exp\left(\int_S \beta(s)ds\right)$ represents the multiplicative change in survival time associated with a one-unit increase in the entire functional covariate $X_i(s)$, holding $\bm{Z}_i$ constant.

\subsection{Model estimation}
Incorporating penalized spline smoothing within a functional modeling framework offers a flexible and computationally tractable approach to analyzing functional data\citep{goldsmith2011penalized, scheipl2015functional}. We adopt this framework to estimate both the \textit{lfAFT} and \textit{afAFT} models. 

To estimate the \textit{lfAFT} model, we begin by expressing the coefficient function as a basis expansion: $\beta(s)=\sum_{k=1}^K b_kB_k(s)$, where $B_1(s),\dots,B_K(s)$ are basis functions such as B-splines and $\textbf{b} = (b_1, \ldots, b_K)^\mathsf{T}$ is a vector of spline coefficients. In practice, the functional covariate for subject $i$ is observed on a discrete, possibly irregular grid $\bm{s}_i = \{s_{i1}, \dots, s_{ip_i}\}$, where the number of observations $p_i$ and the locations $s_{ij}$ may vary across subjects. To approximate the integral in Equation~\ref{eq:2}, we apply numerical quadrature separately for each subject:

\begin{equation}
\begin{aligned}
\log T_i
&= \bm{Z}_i^\mathsf{T}\bm{\gamma} + \int_S \sum_{k=1}^K b_k B_k(s) X_i(s) \, ds + \sigma\epsilon_i \\
&\approx \bm{Z}_i^\mathsf{T}\bm{\gamma} + \sum_{j=1}^{p_i} q_{ij} X_i(s_{ij}) \sum_{k=1}^K b_k B_k(s_{ij}) + \sigma\epsilon_i \\
&= \bm{Z}_i^\mathsf{T}\bm{\gamma} + \sum_{k=1}^K b_k \left\{ \sum_{j=1}^{p_i} q_{ij} X_i(s_{ij}) B_k(s_{ij}) \right\} + \sigma\epsilon_i ,
\end{aligned}
\label{eq:3}
\end{equation}
where $q_{ij}$ are quadrature weights specific to subject $i$. For equally-spaced grid points $s_{ij}$ over $[0,1]$, $q_{ij}=1/p_i$ is sufficient, while for irregular grids, numerical integration rules such as the trapezoidal or Simpson’s rule can be applied to $\bm{s}_i$. The inner sum defines a set of subject-specific covariates: $C_{ik}=\sum_{j=1}^p q_{ij} X_i(s_{ij}) B_k(s_{ij}), \bm{C}_i=(C_{i1},\dots, C_{iK})^\mathsf{T}.$ Thus, the functional AFT model reduces to a standard AFT model with covariates $\bm{Z}_i$ and $\bm{C}_i$, though this will be augmented with penalized smoothing as we describe below.

The distribution of the error term $\epsilon_i$ determines the likelihood. For example, under a log-normal \textit{lfAFT} model, $\epsilon_i \sim N(0,1)$ implies $\log T_i \sim N(\bm{Z}_i^\mathsf{T}\bm{\gamma} + \bm{C}_i^\mathsf{T}\bm{b}, \sigma^2)$. Under a log-logistic \textit{lfAFT} model, $\epsilon_i \sim \text{Logistic}(0,1)$ implies $\log T_i \sim \text{Logistic}(\bm{Z}_i^\mathsf{T} \bm{\gamma} + \bm{C}_i^\mathsf{T} \bm{b}, \sigma)$. This framework can be readily extended to other distributions such as Weibull, depending on the assumed form of $\epsilon_i$. To avoid overfitting and ensure smoothness, a penalty on the curvature of $\beta(s)$ is added to the AFT log-likelihood, resulting in the penalized log-likelihood:

\begin{equation}
\ell(\bm{b},\sigma|\bm{Y},\bm{\delta})=\sum_{i=1}^n\delta_i\log f(Y_i|\bm{b},\sigma)+(1-\delta_i)\log S(Y_i|\bm{b},\sigma)-\lambda \bm{b}^\mathsf{T}\bm{D}\bm{b},
\label{eq:4}
\end{equation}
where $f(\cdot)$ and $S(\cdot)$ are the density function and survival function of $\epsilon_i$, respectively, $\bm{D}$ is a known matrix that penalizes the second-order differences of $\bm{b}$, and $\lambda$ is a nonnegative smoothing parameter. 

Let $\eta_i = \bm{Z}_i^\mathsf{T}\bm{\gamma} + \bm{C}_i^\mathsf{T}\bm{b}$. For the log-normal \textit{lfAFT} model, 
$$f(t_i\mid\bm{b},\sigma)=\frac{1}{t_i\sigma\sqrt{2\pi}}\exp\left(-\frac{\left(\log t_i-\eta_i\right)^2}{2\sigma^2}\right),\quad S(t_i|\bm{b},\sigma)=1-\Phi\left(\frac{\log t_i-\eta_i}{\sigma}\right),$$ where $\Phi(\cdot)$ is the standard normal distribution function.
For the log-logistic \textit{lfAFT} model, let $\alpha_i = \exp(\eta_i)$ and $\kappa = 1/\sigma$. Then 
$$f(t_i \mid \mathbf{b}, \sigma) = \frac{\kappa}{\alpha_i}\, \frac{\left( t_i / \alpha_i \right)^{\kappa - 1}}{ \left( 1 + \left( t_i / \alpha_i \right)^{\kappa} \right)^2 }, \quad S(t_i \mid \mathbf{b}, \sigma) 
= \frac{1}{1 + \left( t_i / \alpha_i \right)^{\kappa}}.$$
Parameter estimation is carried out by maximizing the penalized log-likelihood using the quasi-Newton BFGS algorithm \citep{broyden1970algorithm, fletcher1970algorithm, goldfarb1970algorithm, shanno1970algorithm}, yielding estimates of $\bm{b}$ and $\sigma$.

For the \textit{afAFT} model, the bivariate coefficient surface $F(s, x)$ is represented using tensor product splines: $F(s,x)=\sum_{j=1}^{K_S}\sum_{k=1}^{K_X}b_{jk}B_j(s)B_k(x),$
where $B_j(\cdot)$ and $B_k(\cdot)$ are univariate spline bases defined on the domains of $s$ and $x$, respectively, and ${b_{jk}}$ are spline coefficients \citep{mclean2014functional, cui2021additive}.
Using this representation, the integral term in the \textit{afAFT} model can also be discretized and expressed as a linear combination of transformed covariates, similar to the \textit{lfAFT} case. The resulting model takes the same general form: $\log T_i=\bm{Z}_i^\mathsf{T}\bm{\gamma}+\bm{C}_i^\mathsf{T}\bm{b}+\sigma\epsilon_i$, where $\bm{C}_i$ now consists of subject-specific features constructed from the tensor product basis evaluated at $(s_{ij}, X_i(s_{ij}))$.

\subsection{Optimization of the smoothing parameter}
The smoothing parameter $\lambda$ is selected based on the generalized cross-validation (GCV) criterion. Originally introduced in the context of Ridge regression to approximate leave-one-out cross-validation \citep{golub1979generalized}, GCV has since been broadly applied in spline smoothing and penalized likelihood-based models \citep{ruppert2003semiparametric, gu2013smoothing, wood2017generalized}. In penalized likelihood framework, the GCV score for a given $\lambda$ is defined as:

\begin{equation}
\text{GCV}(\lambda) = -\frac{1}{n} \ell_{np}(\hat{\bm{b}}_\lambda, \hat{\sigma}_\lambda) \bigg/ \left(1 - \frac{\text{df}(\lambda)}{n} \right)^2,
\end{equation}

where $n$ is the number of subjects, $\ell_{np}(\hat{\bm{b}}_\lambda, \hat{\sigma}_\lambda)$ is the unpenalized log-likelihood evaluated at the penalized estimates, and $\text{df}(\lambda)$ is the effective degrees of freedom given by
\begin{equation}
\text{df}(\lambda) = \text{tr}\left(\bm{S}_\lambda\right)
= \text{tr}\left((\bm{C}^\top \bm{W} \bm{C} + \lambda \bm{D})^{-1} \bm{C}^\top \bm{W} \bm{C} \right),
\end{equation}
where $\bm{C}$ is the design matrix built from $\bm{C}_i$, and $\bm{W}$ is a weight matrix derived from the negative second derivative of the unpenalized log-likelihood. A logarithmic grid search over $\lambda \in(1, 10000)$ is conducted to select the value of $\lambda$ that minimizes the GCV score.

\subsection{Inference}
 Pointwise confidence intervals for model parameters are constructed using nonparametric bootstrap with 2{,}000 resamples. Bootstrap-based inference is generally recommended in small-sample settings or when strong smoothing penalties are applied \citep{wood2017generalized}. Alternatively, asymptotic Wald-type 95\% confidence intervals can be obtained based on large-sample normal approximations to the penalized maximum likelihood estimators. In this case, standard errors are derived from the observed information matrix, computed as the Hessian of the penalized log-likelihood with respect to model parameters $(\bm{\gamma}, \bm{b}, \sigma)$, conditional on the selected smoothing parameter $\hat{\lambda}$.

\subsection{Implementation}
The proposed methods are implemented in the R package \texttt{funAFT}, which is publicly available on GitHub. The package supports both linear and additive functional AFT models and allows users to specify the type of smoothing basis, the number of basis functions $K$, and the assumed error distribution (log-normal or log-logistic). Parameter estimation for $\bm{b}$ and $\sigma$ in Equation \ref{eq:4} is carried out using the quasi-Newton BFGS algorithm via the \texttt{optim} function. 

As an alternative, the log-normal \textit{lfAFT} and \textit{afAFT} models can be fit using the \texttt{gam} function from the \texttt{mgcv} package. \citet{crainiceanu2024functional} demonstrates how this software, originally developed for semiparametric regression, can be adapted for survival analysis with functional predictors through functional Cox models. Building on this idea, we extend its use to functional AFT models. To fit these models using \texttt{mgcv}, the log-transformed outcome $\log Y_i$ is modeled as a sum over $s_j$ of a smooth function $\beta_1(s_j)$, weighted by $q_jX_i(s_j)$. While this approach is fast and easily implementable, it is only available for the log-normal family in the current version of software. 

\section{Simulations}
\label{simulations}
\subsection{Simulation design and implementation}
We assess the performance of our methods through simulations designed to mimic the structure of the motivating data. Each simulated dataset contains $n$ subjects, consisting of survival outcome $\bm{Y}$, censoring indicator $\bm{\delta}$, and a functional covariate $\bm{X}(s)$ observed at $p$ evenly spaced points over the domain $[0,1]$. The functional covariate $\bm{X}(s)$ is generated using functional principal components (FPCs) estimated from the motivating pupillometer data. For each subject, new FPC scores are simulated from a multivariate normal distribution with mean zero and covariance matrix equal to the diagonal matrix of eigenvalues, preserving the variance structure of the original data. Survival times $T_i$ are generated under five data-generating models:

\begin{longlist}
\item[1.]
\textit{Linear log-normal AFT model}: $\log T_i=0.5+\int_\mathcal{S}X_i(s)\beta(s)ds+0.5\epsilon_i, $
where $\epsilon_i \sim N(0,1)$ and $\beta(s)$ is a prespecified coefficient function represented by cubic B-splines with six knots and fixed basis coefficients.
\item[2.]
\textit{Linear log-logistic AFT model}: Same as (1), except that $\epsilon_i \sim \text{Logistic}(0,1)$.
\item[3.]
\textit{Linear CoxPH model}: Following \citet{cui2021additive}, a linear functional Cox model (\textit{LFCM}) is fit to the motivating data to estimate the cumulative baseline hazard $\hat{\Lambda}_0(t)$. The linear predictor is defined as $\eta_i=\int_SX_i(s)\beta(s)ds$, with $\beta(s)=0.3-(s-0.2)^2$. Survival times are then simulated from the estimated survival function $\hat{S}_i(t)=\exp\{-e^{\eta_i}\hat{\Lambda}_0(t)\}$.
\item[4.]
\textit{Additive log-normal AFT model}: $\log T_i=0.5+\int_SF[s, X_i(s)]ds+0.5\epsilon_i, $
where $\epsilon_i \sim N(0,1)$ and $F(s,x) = 0.05x^2 s$ is linear in $s$ but nonlinear in $x$.
\item[5.]
\textit{Additive CoxPH model}: Same as (3), except that $\eta_i=\int_SF[s, X_i(s)]ds$ with $F(s,x)=-0.05x^2s$.
\end{longlist}

Censoring times $C_i$ are generated independently from a $\mathrm{Uniform}(0,u)$ distribution to achieve approximately 30\% censoring, with $u = 250$ for linear models and $u = 2000$ for additive models. The observed data are then $Y_i=\min(T_i, C_i)$ and $\delta_i=I(T_i\le C_i)$.

For the linear model scenarios, we generate 500 datasets for each combination of sample sizes $n\in \{100,200,500\}$, grid densities $p\in \{50,100,500\}$, and data-generating models. Each dataset is analyzed using the log-logistic \textit{lfAFT} model, the log-normal \textit{lfAFT} model, the \textit{LFCM} (fitted using \texttt{mgcv}), and the functional time transformation model (FTTM).  Following \cite{ghosal2025functional}, tuning parameters $(N_0, N_1, r)$ for \textit{FTTM} are selected via a grid search minimizing the AIC value. Due to its computational cost, \textit{FTTM} is compared with other methods only in a subset of scenarios ($n\in \{100,200,500\}$, $p = 100$).

For the additive model scenarios, we generate 100 datasets for each combination of sample size $n\in (100,200,500)$, grid density $p = 100$, and data-generating model. Each dataset is fitted with the proposed log-normal \textit{afAFT} and \textit{lfAFT} models, and compared with the additive functional Cox model (\textit{AFCM}, \citet{cui2021additive}) and \textit{LFCM}.

\subsection{Evaluation metrics}
We evaluate the models based on computation time, accuracy of coefficient function and survival function estimation, and predictive performance. For estimation accuracy, we report the mean integrated squared error (MISE), defined as $\text{MISE} = \mathbb{E} \int \{\hat{g}(t) - g(t)\}^2 dt$, for both the estimated coefficient function $\hat{\beta}(s)$ and the subject-specific survival function $\hat{S}_i(t)$, where $g(t)$ denotes the true $\beta(s)$ or $S_i(t)$. Predictive accuracy is assessed using the out-of-sample Brier score, calculated as $\text{Brier score} = \frac{1}{n} \sum_{i=1}^n (p_i - o_i)^2$, where $p_i$ is the predicted survival probability and $o_i$ the observed outcome (1 for event, 0 for non-event). The test dataset is generated independently and has the same sample size $n$ as the training dataset. Both the MISE of $\hat{S}_i(t)$ and the Brier score are computed over the time interval $t\in[0,120]$ minutes and averaged across subjects.

\subsection{Simulation results}
Table~\ref{table1} reports the mean computation times (in seconds) for the log-logistic and log-normal \textit{lfAFT}, log-normal \textit{afAFT}, \textit{LFCM}, \textit{AFCM}, and \textit{FTTM} across eight simulation scenarios. Under the linear model settings, the mgcv-fitted \textit{LFCM} is the most computationally efficient, typically completing within 0.1 seconds per dataset. The log-normal and log-logistic \textit{lfAFT} require slightly longer (1-9 seconds depending on sample size), but still achieve practical runtimes. By contrast, \textit{FTTM} is orders of magnitude slower, taking several hundred seconds per dataset even with a small-scale grid search. Under the additive model scenarios, computation times increase substantially for both the log-normal \textit{afAFT} model and \textit{AFCM}, with mean runtimes ranging from 16 to 28 seconds, but remain comparable between the two methods.


\begin{table}[H]
\caption{Mean computation time (seconds) for model fitting across data-generating models and sample sizes ($n$), with grid density $p = 100$.}
\label{table1}
\centering
\begin{tabular}{@{}lcccccccc@{}}
\hline
\makecell[l]{Data-generating model} & 
\makecell{$n$} & 
\makecell{log-logistic\\lfAFT} & 
\makecell{log-normal\\lfAFT} & 
\makecell{log-normal\\afAFT} & 
\makecell{LFCM} & 
\makecell{AFCM} & 
FTTM \\
\hline
linear log-normal AFT    & 100  & 1.722  & 3.057 & - & 0.030  & - & 285.849  \\
linear log-normal AFT   & 200  & 2.718  & 4.209 & - & 0.046  & - & 440.651 \\
linear log-normal AFT    & 500  & 5.060  & 8.549  & - & 0.092 & - & 896.851 \\
linear log-logistic AFT  & 100  & 1.622  & 2.727  & - & 0.029 & - & 241.088 \\
linear CoxPH        & 100  & 2.409  & 2.877  & - & 0.027 & - & 928.518 \\
additive log-normal AFT    & 100  & - & - & 20.895  & - & 16.606  & - \\
additive log-normal AFT  & 200  & - & - & 21.240 & - & 18.658  & - \\
additive log-normal AFT  & 500  & - & - & 27.512 & - & 28.427  & - \\
\hline
\end{tabular}
\end{table}

\begin{wrapfigure}{r}{0.6\textwidth}
  \centering
  \vspace{-0.5\baselineskip} 
  \includegraphics[width=\linewidth]{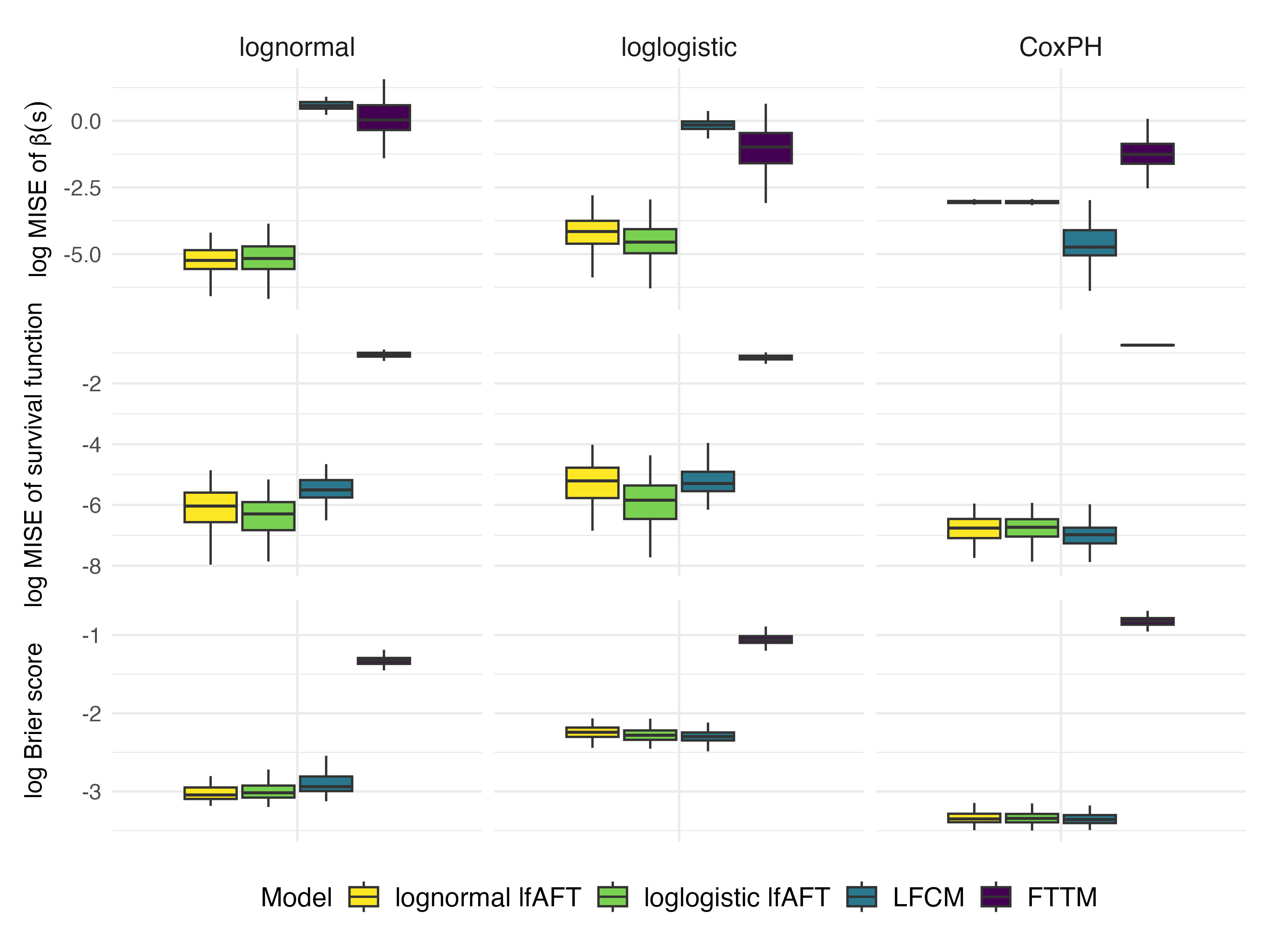}
  \caption{Log MISE of estimated coefficient functions $\hat{\beta}(s)$ (top row), log MISE of estimated survival functions $\hat{S}(t)$ (middle row), and log out-of-sample Brier scores (bottom row) across data-generating models. Results are shown for the subset including \textit{FTTM} with sample size $n=100$ and grid density $p=100$.}
  \label{fig_sim_fttm}
\end{wrapfigure}

Figure~\ref{fig_sim_fttm} compares performance of the proposed log-logistic and log-normal \textit{lfAFT} models with \textit{LFCM} and \textit{FTTM} in a subset of linear model scenarios ($n=100$, $p=100$). For coefficient function estimation (top row), the log-logistic, log-lognormal \textit{lfAFT}, and LFCM each achieve the lowest MISE under their respective data-generating settings, with only marginal difference between the two \textit{lfAFT} models. \textit{FTTM} performs slightly better than \textit{LFCM} in the AFT settings but is clearly outperformed by both \textit{lfAFT} models, and it is inferior to all methods when the true model is CoxPH. For survival function estimation (middle row), log-logistic \textit{lfAFT} attins the highest accuracy in the AFT scenarios and both \textit{lfAFT} models perform comparably to \textit{LFCM} under CoxPH, whereas \textit{FTTM} again shows lower accuracy. In terms of out-of-sample prediction (bottom row) as assessed by Brier score, the \textit{lfAFT} models perform comparably to \textit{LFCM} across all data-generating models, while \textit{FTTM} remains the least accurate. Overall, the proposed \textit{lfAFT} models provide robust estimation and prediction, consistently outperforming \textit{FTTM} and performing comparably to or better than \textit{LFCM}.

\begin{figure}[H]
\includegraphics[width=1\textwidth]{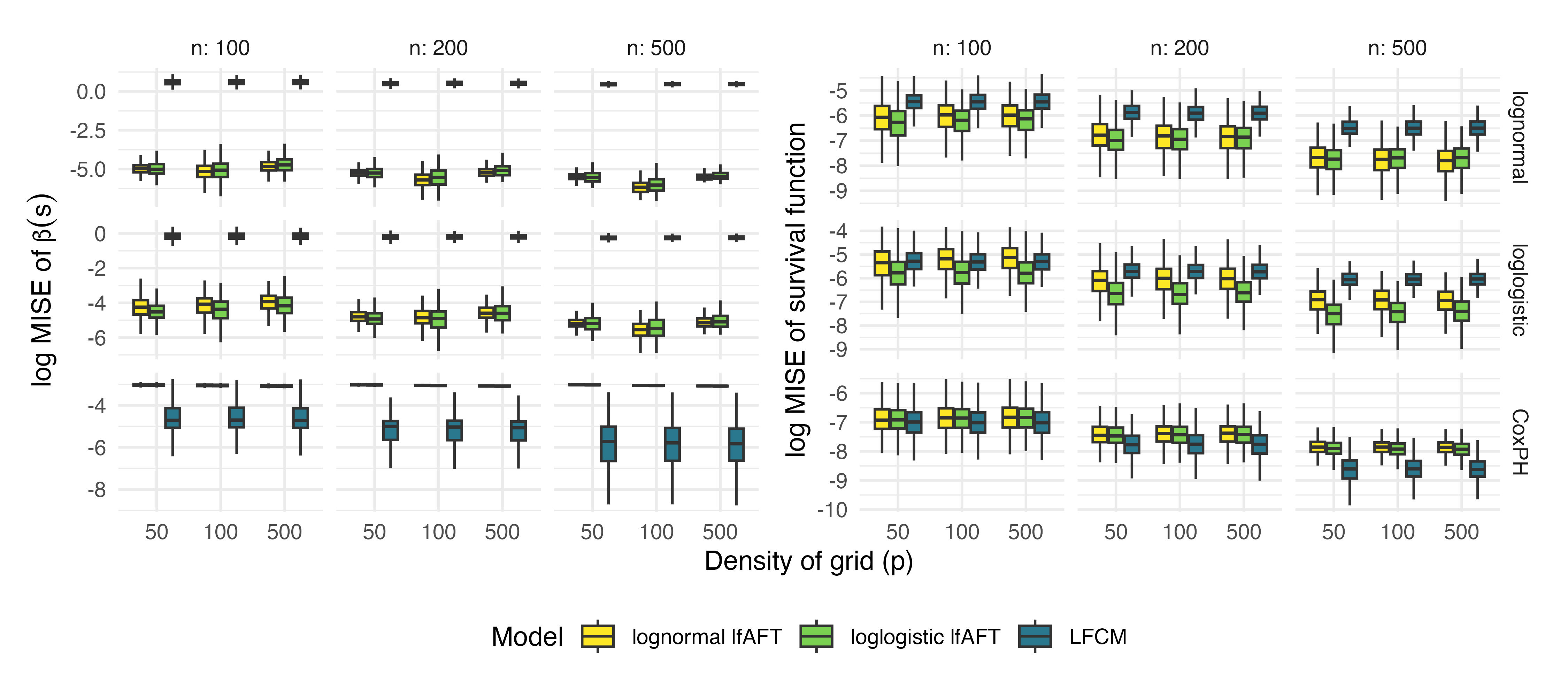}
\caption{Log MISE of estimated coefficient functions $\hat{\beta}(s)$ (left panel) and estimated survival functions $\hat{S}(t)$ (right panel) across data-generating models, sample sizes $(n)$, and grid densities $(p)$.}
\label{fig_sim_mise}
\end{figure}

Figure~\ref{fig_sim_mise} presents the log MISE of estimated coefficient functions $\hat{\beta}(s)$ and survival functions $\hat{S}(t)$ across all linear simulation scenarios, varying data-generating models (log-normal, log-logistic, CoxPH), sample sizes ($n = 100, 200, 500$), and grid densities ($p = 50, 100, 500$). For coefficient function estimation, the log-normal and log-logistic \textit{lfAFT} models yield similarly low MISE in the AFT settings, accurately recovering the true coefficient function even under mild misspecification. When data are generated from the CoxPH model, \textit{LFCM} is the correctly specified model and achieves the lowest MISE for $\hat{\beta}(s)$ as expected. 

For survival function estimation, when sample size is small ($n = 100$), the log-logistic \textit{lfAFT} achieves the best overall performance under the AFT settings and both \textit{lfAFT} models perform comparably to \textit{LFCM} even when the true model is CoxPH. In comparison, \textit{LFCM} exhibits lower accuracy under AFT settings, reflecting sensitivity to model misspecification. Across all scenarios, increasing sample size consistently improves estimation accuracy, while changes in grid density has smaller and less consistent effects.

\begin{figure}[H]
\includegraphics[width=0.8\textwidth]{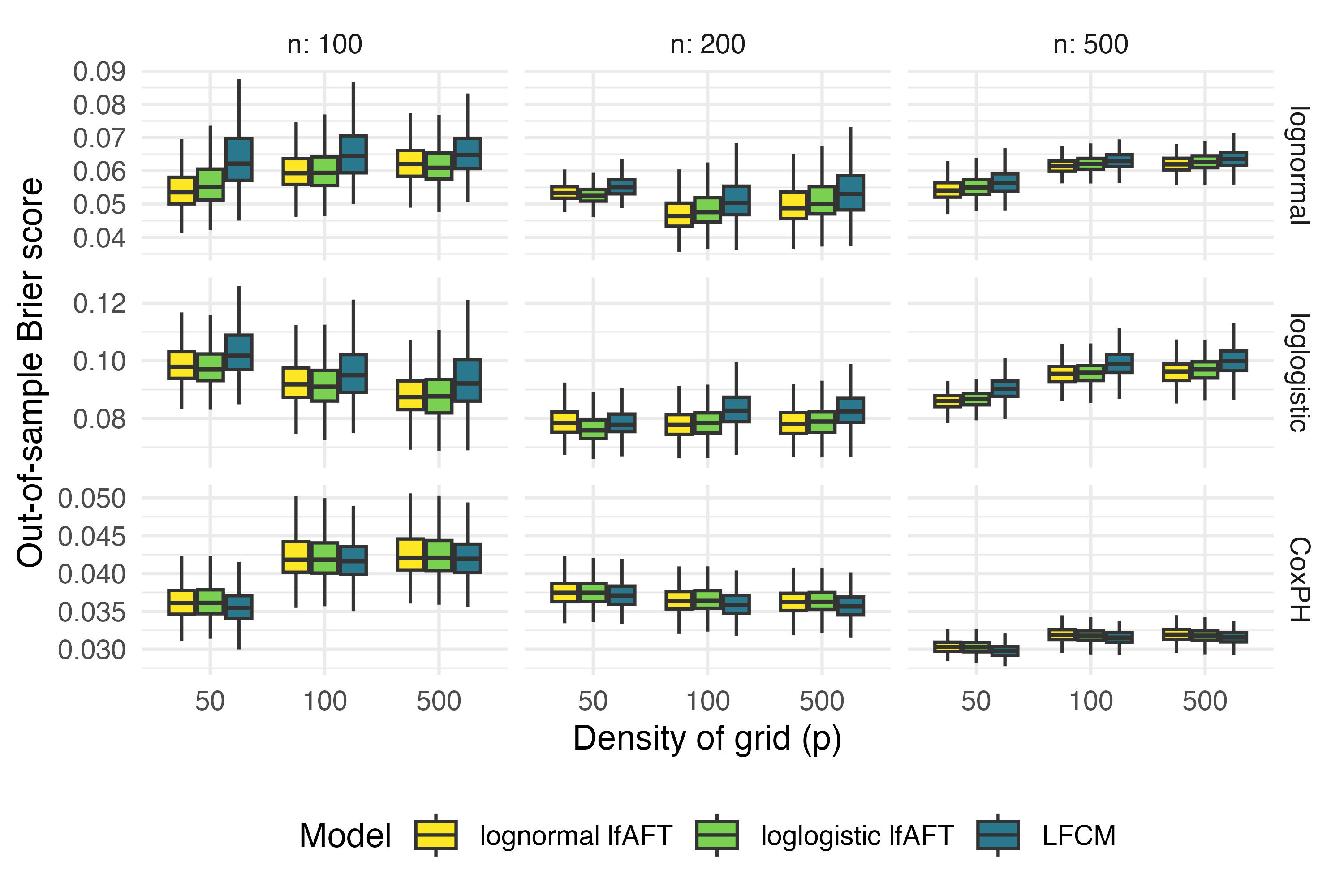}
\caption{Out-of-sample Brier scores across data-generating models, sample sizes $(n)$, and grid densities $(p)$.}
\label{fig_sim_brier}
\end{figure}

Figure~\ref{fig_sim_brier} shows the out-of-sample Brier scores across all linear simulation scenarios. In the AFT settings, the log-normal and log-logistic \textit{lfAFT} achieve comparably low prediction errors across sample sizes and grid densities, while the \textit{LFCM} exhibits slightly higher Brier scores. When data are generated from the CoxPH model, all three models remain close in predictive performance. Across all scenarios, larger sample sizes lead to improved predictive accuracy and reduced variability, while grid density has little impact. Taken together with the results in Figure~\ref{fig_sim_mise}, the proposed \textit{lfAFT} models provide strong and robust predictive performance, matching or exceeding \textit{LFCM} under a range of data-generating mechanisms.

\begin{figure}[H]
\includegraphics[width=1\textwidth]{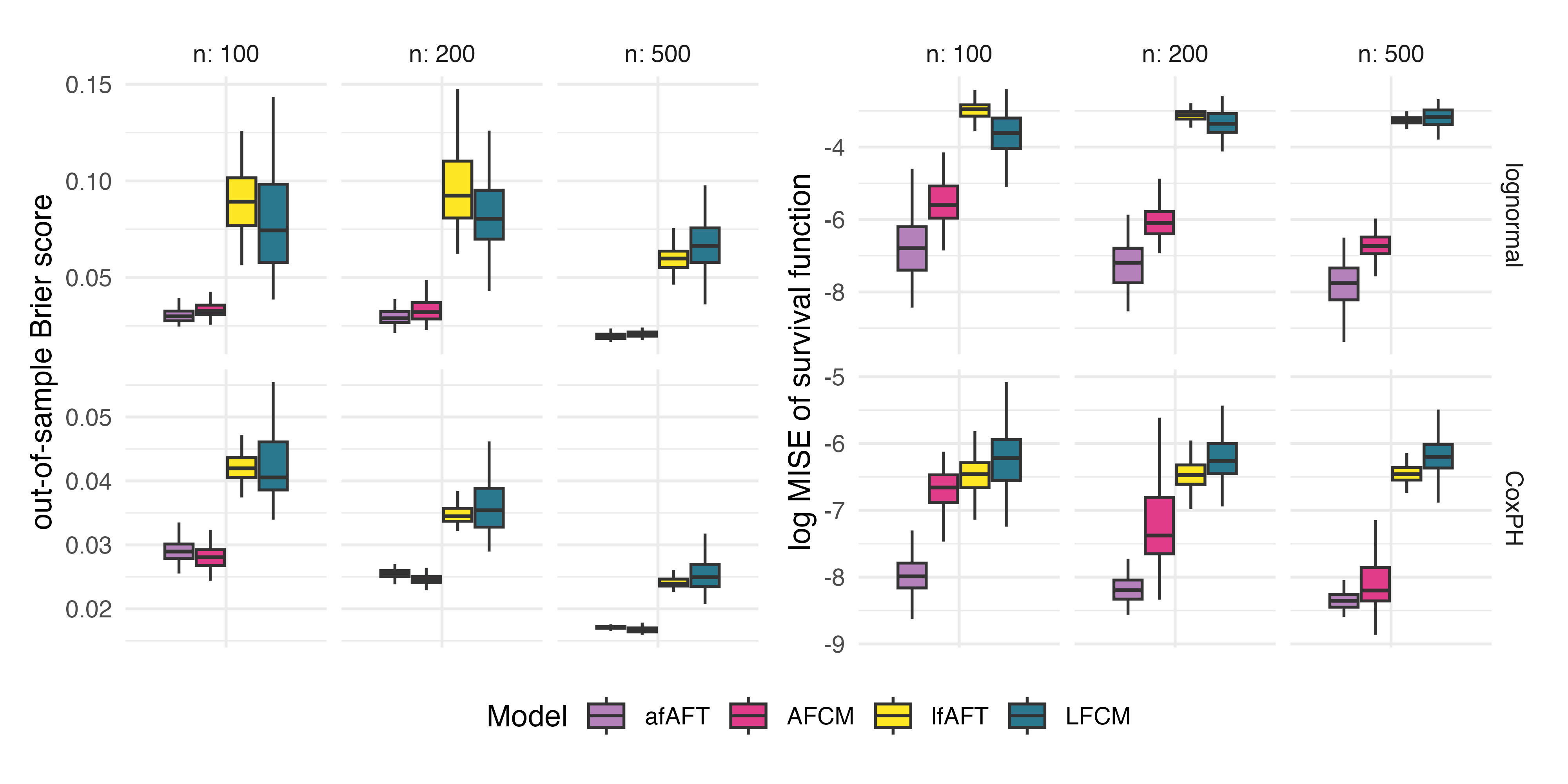}
\caption{Out-of-sample Brier scores (left panel) and log MISE of estimated  survival functions $\hat{S}(t)$ (right panel) across data generating models and sample sizes $(n)$. Results are shown for the additive model scenarios with grid density $p=100$.}
\label{fig_sim_additive}
\end{figure}

Figure~\ref{fig_sim_additive} summarizes the performance of four models under additive simulation scenarios with grid density $p=100$. For out-of-sample prediction (left panels), both \textit{afAFT} and \textit{AFCM} achieve the lowest Brier scores under their respective data-generating settings. In contrast, \textit{lfAFT} and \textit{LFCM} perform poorly in these nonlinear settings, with higher prediction error and greater variability. For survival function estimation (right panels), \textit{afAFT} consistently produces the lowest log MISE across both AFT and CoxPH data-generating models. \textit{AFCM} is less accurate, and lfAFT and \textit{LFCM} underperform due to their restrictive linearity assumptions. These results highlight the advantages of \textit{afAFT} in capturing nonlinear functional effects, yielding more accurate estimation and prediction than existing methods.

\section{Data Analysis}
\label{data analysis}

We now analyze the pupillometer data introduced in Section \ref{StudyDesign}. This dataset includes 127 participants, classified as cannabis users (N = 96) or non-users (N = 31) based on self-reported use history. Users were instructed to smoke ``the amount they commonly use for the effect they most commonly desire'',  while non-users rested for the same duration without smoking. Pupillary dynamics were measured using a pupillometer at approximately 40 and 100 minutes after the start of cannabis inhalation. At each assessment, pupil diameter was recorded at 30 Hertz for about four seconds following a brief light stimulus (180 microwatts), resulting in a grid shared across participants of 120 measurements per curve. To reduce variability due to baseline pupil size, we compute percent change from the start of the test and use this trajectory as the functional predictor $X_i(s)$. Analyses are restricted to the right eye; given the strong correlation between eyes, similar results are expected for the left.

The outcome $T_i$ is defined as time since cannabis use, in minutes. Separate models are fit for the first assessment (median = 40 minutes, range = 27-56) and the second assessment (median = 97 minutes, range = 85-120). For non-users, $T_i$ is censored at 480 minutes, as participants were asked to abstain from cannabis use for at least eight hours prior to their visit; we are treating this as a right censoring mechanism. All models include the functional covariate $X_i(s)$ and scalar covariates age and BMI.

For each assessment, we apply the log-logistic and log-normal \textit{lfAFT} models, and log-normal \textit{afAFT} model as described in Section \ref{methods}, and interpret the estimated parameters from each model in the context of the motivating experiment. The \textit{lfAFT} models use a B-spline basis with 20 basis functions, while the \textit{afAFT} employs a tensor product basis with $20 \times 20$ basis functions. Predictive performance is evaluated via ten-fold cross-validation, summarizing Brier scores over a follow-up window of $\tau$ minutes (60 for Assessment 1, 120 for Assessment 2). As a benchmark, we compare these results with a functional logistic regression model that predicts the binary outcome of whether an individual consumed cannabis within the last $\tau$ minutes. In terms of computation, the log-logistic \textit{lfAFT} requires 1.98 seconds per assessment period, the log-normal \textit{lfAFT} 0.89 seconds, and the log-normal \textit{afAFT} 27.10 seconds.

\subsection{Estimated coefficient function and surface}
\begin{figure}[H]
\includegraphics[width=1\textwidth]{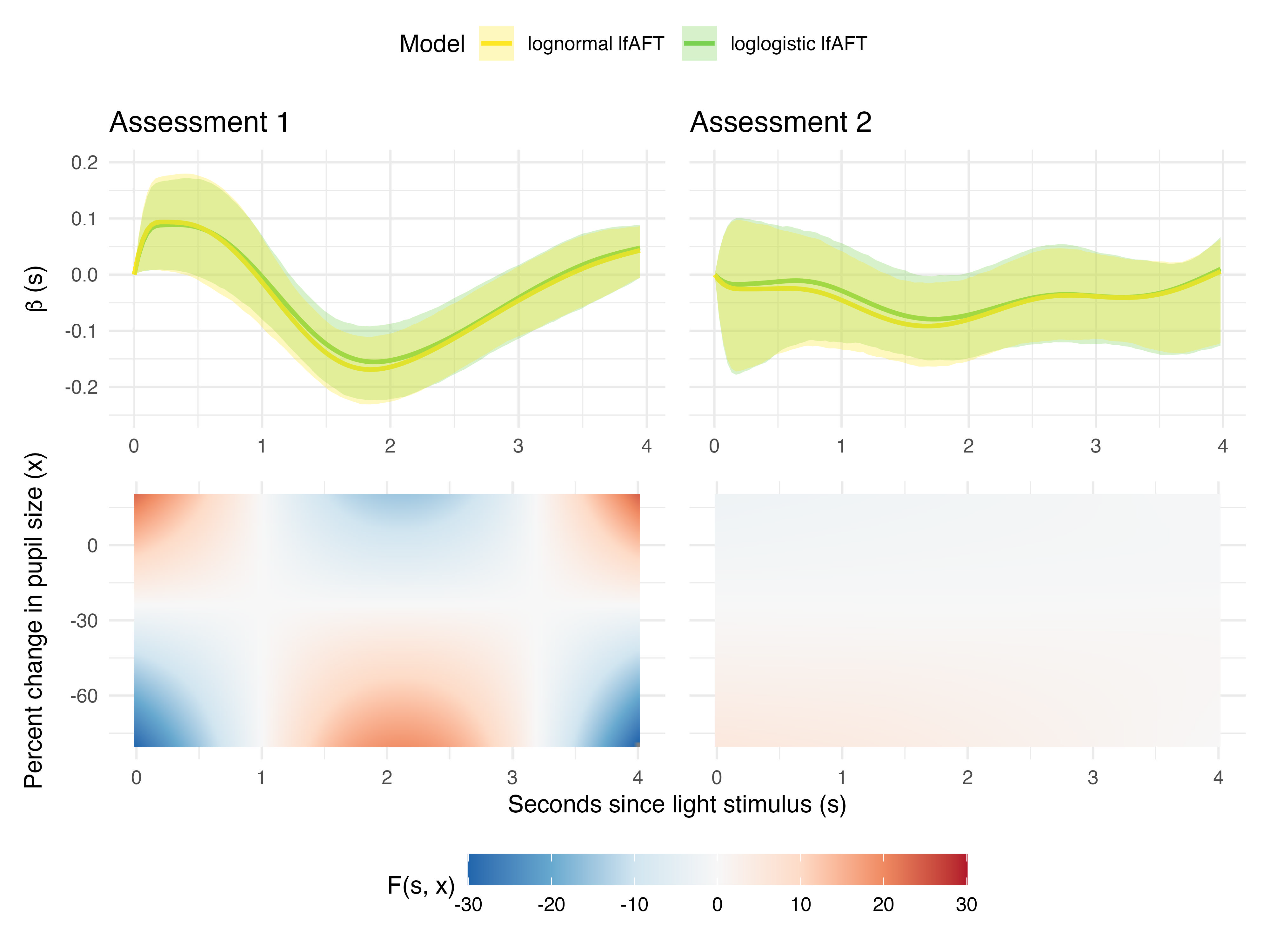}
\caption{Estimated coefficient functions, $\hat{\beta}(s)$, with 95\% pointwise bootstrap confidence intervals from the \textit{lfAFT} models (top row) and estimated coefficient surfaces, $\hat{F}(s,x)$, from the log-normal \textit{afAFT} model (bottom row) for pupil light response data at Assessment 1 (left) and Assessment 2 (right).}
\label{fig_data_coef}
\end{figure}

The upper panels of Figure~\ref{fig_data_coef} display the estimated coefficient functions, $\hat{\beta}(s)$, with 95\% pointwise bootstrap confidence intervals (CIs) from the log-normal and log-logistic \textit{lfAFT} models. At Assessment 1 (approximately 40 minutes post-smoking), both models reveal a U-shaped relationship between pupil light response and time since cannabis use. In the log-logistic model, $\hat{\beta}(s)$ is significantly negative between 1.239 and 2.844 seconds and reaches its largest absolute value at 1.874 seconds, which corresponds to the point of minimum constriction (Figure \ref{fig_curves}). This pattern indicates that smaller percent change from baseline to minimum pupil size (e.g. reduced pupil constriction) is associated with shorter time since smoking, reflecting more recent use. The exponential of the area under the curve, $\exp\{\int_\mathcal{S}\hat{\beta}(s)ds\}\approx1.078$, represents the multiplicative change in time since cannabis use associated with a one-unit (percent change) increase in the entire pupil light response curve. The estimated scale parameter $\sigma$ is 0.700 (95\% CI: 0.573–0.826), consistent with a unimodal hazard function for recent smoking. The log-normal model produced nearly identical coefficient estimates and CIs, with $\hat{\sigma}= 1.264$, corresponding to the standard deviation of log survival time around the linear predictor. At Assessment 2 (approximately 100 minutes post-smoking), both models yield estimates close to null, with $\hat{\beta}(s)$ significantly negative only between 1.373 and 2.041 seconds in the log-normal model, again corresponding to the point of minimum constriction. This suggests that the predictive signal of pupil dynamics diminishes as the time since cannabis use increases and impairment subsides.

The lower panels of Figure~\ref{fig_data_coef} show the estimated coefficient surfaces, $\hat{F}(s,x)$, from the log-normal \textit{afAFT} model. At Assessment 1, the surface reveals a nonlinear interaction between the timing of the pupil response and the magnitude of percent change in pupil size. During the light onset (0–1 second), an initial constriction greater than 30\% is associated with longer time since smoking, whereas smaller constriction or dilation (less than 30\%) is associated with more recent use. Around the point of minimal constriction and early rebound (1.4–2.7 seconds), the estimated effects change direction: stronger constriction (greater than 30\%) after 1.4 seconds is linked to longer time since smoking, while weaker constriction or rebound is linked to shorter times. At Assessment 2, the estimated surface is nearly flat, suggesting little to no association between pupil dynamics and time since cannabis use.

\subsection{Predictive performance}
Table \ref{table2} reports the 10-fold cross-validated Brier scores from the log-logistic and log-normal \textit{lfAFT} models, the log-normal \textit{afAFT} model, and functional logistic regression. For both assessments, the \textit{lfAFT} and \textit{afAFT} models achieve nearly identical Brier scores (0.1270-0.1286 at Assessment 1, 0.0982-0.1011 at Assessment 2), suggesting that the additive \textit{afAFT} framework offers no additional predictive benefit in this setting and that the simpler \textit{lfAFT} model is sufficient. The log-logistic and log-normal \textit{lfAFT} models also perform comparably, with the former showing a marginal advantage, indicating that choice of error distribution does not substantially impact prediction in this context. In contrast, functional logistic regression shows inferior predictive accuracy (0.1736 at Assessment 1, 0.1721 at Assessment 2), highlighting the advantage of modeling time since cannabis use as a continuous survival outcome rather than reducing it to a binary classification problem.

\begin{table}[H]
\caption{10-fold cross-validated Brier scores across models and assessments.}
\label{table2}
\centering
\begin{tabular}{@{}lcc@{}}
\hline
\makecell[l]{Model} & 
\makecell{Brier Score (Assessment 1)} & 
\makecell{Brier Score (Assessment 2)} 
\\
\hline
log-logistic \textit{lfAFT}     & 0.1275  & 0.0982 \\
log-normal \textit{lfAFT}       & 0.1286  & 0.1001 \\
log-normal \textit{afAFT}       & 0.1270  & 0.1011 \\
functional logistic   & 0.1736  & 0.1721 \\
\hline
\end{tabular}
\end{table}

\section{Discussion}
\label{discussion}

In this work, we propose two new models for analyzing right-censored survival data with functional covariates: the additive functional accelerated failure time model (\textit{afAFT}) and its special case, the linear functional accelerated failure time model (\textit{lfAFT}). We demonstrate their utility in predicting time since cannabis use from pupil light response curves, a setting where identifying objective biomarkers of recent use is of significant public health and traffic safety importance.

The \textit{afAFT} model extends the traditional linear functional approach by allowing the effect of the functional covariate to vary smoothly with both time within the functional domain and the magnitude of the covariate. This flexibility enables \textit{afAFT} to capture nonlinear and interactive effects that may be obscured under a linear assumption. For the pupil response data, this has important practical implications: changes in pupil diameter during initial constriction and subsequent rebound dilation may convey distinct information about time since smoking, and \textit{afAFT} provides a framework to quantify these potentially heterogeneous effects.

Our simulation study shows that the proposed models provide accurate estimation of coefficient and survival functions and deliver strong predictive performance relative to existing methods, including the linear and additive functional Cox models (\textit{LFCM}/\textit{AFCM}) and the functional time transformation model (\textit{FTTM}). The \textit{lfAFT} is robust to moderate model misspecification, and the \textit{afAFT} excels in nonlinear settings. In addition, both models are computationally efficient, straightforward to implement in R, and supported by publicly available reproducible code, which facilitates practical adoption in applied research.

When applied to the motivating pupillometer data, our models provide novel insights into the relationship between pupil dynamics and cannabis use. At the 40-minute post-smoking assessment, the \textit{lfAFT} identifies that greater percent change from baseline to minimum pupil size is associated with longer time since smoking. The \textit{afAFT} further reveals a nonlinear interaction between the timing of the pupil response and the magnitude of percent change. At the 100-minute assessment, both models show little or no association, suggesting that the predictive signal of pupil dynamics diminishes over time as acute cannabis effects subside. These findings align with prior evidence that cannabis-related impairment is strongest 20–40 minutes following inhalation \citep{sewell2009effect}.

Our study has several limitations. First, we fit separate models for each assessment, thereby ignoring the repeated-measures structure of the data. Extending the current framework to accommodate longitudinal survival outcomes with functional predictors would better exploit within-subject correlation. Second, while we focus on parametric AFT models with log-normal and log-logistic errors, future work could consider semiparametric functional AFT models that allow flexible error distributions yet accomodate the flexible additive structure of the predictors. Finally, while our motivating application focuses on pupil light response, the proposed methods are broadly applicable to other domains where functional covariates predict time-to-event outcomes, such as wearable health monitoring or neuroimaging studies.

\section*{Acknowledgments}
This research was supported by the National Institute on Drug Abuse (NIDA) of the National Institutes of Health (NIH) under R01DA049800 and P50DA056408, and by the Emory University Rollins School of Public Health Dean’s Pilot Innovation Award.

\newpage

\bibliographystyle{imsart-nameyear} 
\bibliography{bibliography}       

\end{document}